\documentclass[preprint,aps,pra,amsmath,amssymb,superscriptaddress,floatfix]{revtex4-2}

\usepackage{graphicx}
\usepackage{bm}
\graphicspath{{figs/}}
\usepackage{xcolor}%

\begin{document}


\title{Nonperturbative theory of thin-vapor-layer linear optical properties: towards the optimal design of alkali-metal vapor nanocells}

\author{A. V. Ermolaev}\homepage{Corresponding author. Email: ermolaevandrej1997@gmail.com. https://orcid.org/0000-0001-6144-5124}
\affiliation{Independent researcher, Bordeaux, France}
\author{T. A. Vartanyan}\homepage{https://orcid.org/0000-0002-3956-9232}
\affiliation{ITMO University, Kronversky Prospect 49, 197198 St. Petersburg, Russia} 

\begin{abstract}
Hot alkali-metal vapors confined inside nanometer-thick spectroscopic cells are
highly suitable media for miniature and cheap frequency standards, magnetometers and
atomic sensors. Yet their linear optical properties at low to moderate atomic densities are not fully understood. The main reason is the complex,
strongly nonlocal optical response of such systems to resonant radiation, which arises from the free motion of atoms and their collisions with the walls. To date, calculations of reflection and transmission
spectra of these cells have been restricted to perturbation theory in the atomic
density for the case of quenching atom-wall collisions, which has substantially limited their practical use. Here we introduce a
rigorous numerical approach for computing thin-vapor-layer spectra that accurately accounts for the nonlocal optical response at
arbitrary optical density and layer thickness. Our model predicts a density-dependent
blue shift of the resonance that is absent in the first-order perturbative description. We show that its values are comparable to the homogeneous width of an atomic transition and, therefore, too large to be neglected in most cases.
Next, inspired by recent experiments, we explore a wide range of vapor layer thicknesses and study how the reflectivity and its derivative with respect to the laser detuning depend on the layer thickness. We obtain an approximate expression for the thickness maximising the first derivative of the reflectivity that
agrees closely with the rigorous numerical model and reduces to the first-order analytical result showing the optima at $l=\lambda/3+n\lambda$ and $l=2\lambda/3+n\lambda$, where $n$ is a non-negative integer. These results provide a
practical design tool for nanometer-thick spectroscopic cells in various applications and a starting point for further
development of the theory.
\end{abstract}

\maketitle

\newpage

\section{Introduction}
\label{sec:intro}

Alkali-metal vapors confined between two dielectric windows in a layer as thick as a fraction of a
wavelength are now a standard platform for miniature frequency
references, magnetometers, clocks, and chip-scale photonic
devices~\cite{KITCHING,APP1,APP2,CHIP,Ritter,URIEL:21,LEVY,Uriel}. They can also serve as a testbed for exploring fundamental aspects of atom--surface interactions, collisional processes and the associated line
shifts and
broadenings~\cite{ADAMS,ADAMS2,KEAVENEY,PhysRevA.97.053841,PhysRevLett.123.173203,PhysRevApplied.14.034054}.
Since the first sub-micron cells became available~\cite{SARKISYAN2001}, they have also
been used for detailed studies of the hyperfine
structure of alkali
metals and various optical and quantum effects~\cite{Sargsyan:12,Sargsyan:2014,Sargsyan:17,Sargsyan:17_2,Sargsyan2026}. 

Doppler-free
structures in the spectra of light reflection from a dielectric--vapor interface were first observed by Cojan in mercury~\cite{Cojan1954}. Having demonstrated that the conventional dispersion theory, which omits atom-wall collisions, fails to explain the observed phenomenon, he developed a microscopic theory that accounts not only for the motion of atoms, but also for their collisions with the walls of the cell. The case of specular rebound with the preservation of polarisation was examined in the greatest detail. Just after that, Romer and Dicke~\cite{DICKE} discovered the disappearance of Doppler broadening in radio-frequency radiation of ammonia placed in a gas cell consisting of two flat parallel walls, separated by approximately half a wavelength. Later, Woerdman and co-workers~\cite{WOERDMAN1975248,burgmans:jpa-00208462} observed Doppler-free reflection spectra in sodium vapor and Schuurmans derived an exact solution for the semi-infinite vapor layer problem~\cite{schuurmans:jpa-00208442}. It was shown that the possibility of observing Doppler-free resonances is due to the strong nonlocality of optical response of such vapor layers at low and moderate atomic densities. Specifically, the physical
origin of this effect was attributed to the transient polarisation component that contributes significantly only when the mean free path of an atom becomes comparable to the wavelength of the incident radiation. This regime can be achieved at low to moderate atomic concentrations, in which case the thermal motion of atoms and their collisions with boundaries
together lead to the appearance of sub-Doppler spectral contours in the linear regime of interactions with the incident field.

Work on
selective reflection has since covered nonlinear
processes~\cite{VART1985,DUCLOY1988}, laser-line narrowing~\cite{VART1990},
high-resolution spectroscopy~\cite{SINGH1986107,Papoyan1,Papoyan2},
higher-order density effects~\cite{GUO1994}, antireflection
coatings~\cite{VARTANYAN1994} and the Lorentz--Lorenz
correction~\cite{GUO1996}, and identified a density-sensitive blue shift of the
resonance~\cite{BLUESHIFT} whose origin remained
unexplained~\cite{GUO1996}. Later, it was predicted that the nonlocal optical response can be enhanced if the vapor is confined in a layer with thickness of the order of the incident wavelength~\cite{PhysRevA.51.1959,ZAMBON1997308}. 
Despite this progress, several important limitations remained. 
The early treatment of the thin-vapor-layer problem
neglected the reflection from the rear window and used the first-order perturbation theory (PT) in atomic density. Dutier and co-workers later removed
the first limitation by mixing the atomic response with the Fabry--P\'erot (FP) cavity
contribution~\cite{Dutier:03}, bringing the calculated
spectra considerably closer to the measured ones, but their description remained
first order in atomic density. 

An exact solution, free of any expansion in the atomic
density, was later obtained in Ref.~\cite{PhysRevA.101.053850}, although under
the assumption that atoms are reflected specularly with their polarisation
preserved. The physically better motivated assumption, in which the
polarisation is quenched after atom-wall collisions~\cite{Zajonc,Duclou1993,Todorov,Todorovproc}, was treated in
Ref.~\cite{EV2022} with the solution built again through the expansion in the atomic
density, although extended to higher orders. PT restricts the analysis to optically-thin layers
where the expansion converges, leaving cells of intermediate length and/or vapors at higher densities without any
quantitative description. Mathematically speaking, the described system for the case of quenching collisions can be treated as a
boundary-value problem with a nonlocal relation between the field and polarisation, of the type met
in metallic slabs with a finite electron mean free path~\cite{Baraff}. However, the proposed analytical Wiener-Hopf method leads to very cumbersome solutions. 

In this work, we remove
the second limitation. Specifically, for quenching atom-wall collisions we reduce the self-consistent problem to a
single integral equation for the field and solve it exactly by direct matrix
inversion. To our knowledge, this is the first rigorous numerical scheme for the considered thin-vapor-layer
problem. It is accurate at any optical density and at any thickness, and so
describes sub-wavelength layers and cells of intermediate length. This matters because many recent studies  actively explore the selective reflection from a nanocell~\cite{Sargsyan:12,Sargsyan:2014,Sargsyan:17,Sargsyan:17_2,Sargsyan2026}, however without proper optimisation of the vapor layer thickness, which is known to be one of the parameters that most strongly influences the observed spectral line shape.

We also study in
detail how the amplitude of the first derivative of the reflectivity with respect to the laser detuning~\cite{DucloyFichet1991} depends on
the layer thickness, and show that this dependence is a complex function of
the window reflectivity. We
derive an approximate analytical expression for the thickness that maximises the amplitude of the first derivative of reflectivity with respect to detuning and show that for
surrounding media of low refractive indices the optimum cell
thicknesses correspond to $\lambda/3 + n\lambda$ and $2\lambda/3 + n\lambda$, where $n= 0,1,2,...$. Finally, we demonstrate that the reflection spectra computed accurately exhibit an additional shift in the resonance frequency when compared to the results of first-order PT. The magnitude of this shift is shown to be of the same order as the homogeneous linewidth.

The paper is organized as follows. Section~\ref{sec:methods} presents the geometry
and notation and recalls the main assumptions used in theoretical description. In Sec.~\ref{sec:results} we present the exact solution of the described thin-vapor layer problem along with the implemented numerical scheme. We next derive an approximate analytical expression for the optimal layer thickness maximising the amplitude of the first derivative of the reflectivity with respect to laser detuning. 
In Sec.~\ref{sec:discussion} we
discuss the obtained results and conclude. Finally, Appendices~\ref{app:log} and~\ref{app:extra} provide additional information needed for the derivation of the optimal layer thickness and examples of calculated reflection spectra.

\section{Theoretical framework}
\label{sec:methods}


\subsection{Geometry, assumptions and notation}
\label{sec:setting}

\begin{figure}[h]
\centering
\includegraphics[width=0.5\linewidth]{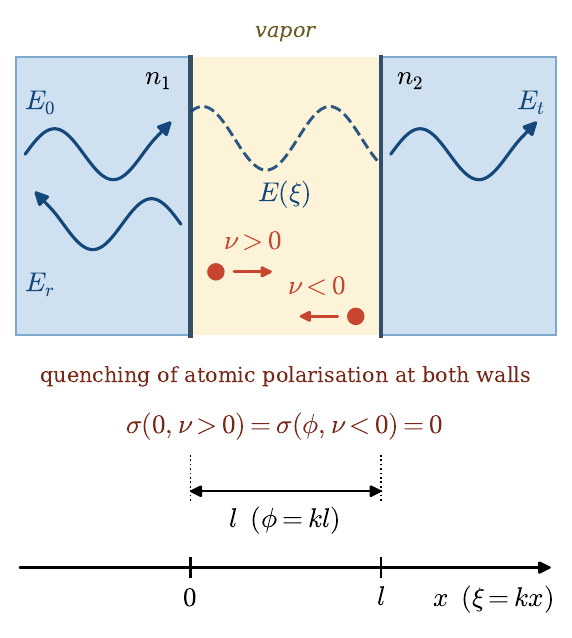}
\caption{\label{fig:geometry}Geometry of the problem. A resonant vapor occupies
the region $0\leq x\leq l$ between two semi-infinite transparent dielectrics of
refractive indices $n_{1}$ and $n_{2}$. The incident, reflected and transmitted
amplitudes are $E_{0}$, $E_{r}$ and $E_{t}$, and $E(\xi)$ is the
unknown field inside the layer. Atoms are considered to
leave either wall with no polarisation.}
\end{figure}

We consider resonant light interaction with an atomic vapor confined between two
semi-infinite transparent dielectric media. The outer boundaries of
these media play no role and are omitted. The vapor consists of two-level atoms with transition frequency $\omega_{21}$.
The normally incident field is a linearly polarized, monochromatic plane
wave of frequency $\omega$ tuned close to the resonance frequency (i.e. $|\omega-\omega_{21}|/\omega
\ll1$), so that the rotating-wave approximation applies. The incident power is low
enough not to saturate the transition and optical pumping is neglected. Within these assumptions the vapor response remains linear with respect to the incident field, while
the strong spatial dispersion induced by thermal motion is retained in
full~\cite{VART2001,PhysRevA.101.053850}. We note here that the atoms' velocity components perpendicular to the direction of light propagation do not result in a Doppler shift and can be disregarded. The only component of the atomic velocity that is relevant for the treatment is the projection onto the direction of the beam $\upsilon$ ($x$-axis in Fig.~\ref{fig:geometry}). Therefore, we assume that the distribution of atoms along this velocity projection corresponds to a one-dimensional Maxwell distribution~\cite{Todorov,Todorovproc,LIFSHITZ1981329}, which reads as $f(\upsilon) = (\sqrt{\pi}\upsilon_T)^{-1}\exp{(-\upsilon^2/\upsilon_T^2)}$ with $\upsilon_T$ denoting the most probable atomic velocity at a given temperature of the atomic vapor $T$. The described one-dimensional geometry of the problem is shown
in Fig.~\ref{fig:geometry}, where $l$ is the thickness of the gas layer and
$n_{1}$, $n_{2}$ are the refractive indices of the surrounding media.

Following Ref.~\cite{EV2022}, we use the dimensionless variables
\begin{equation}
\begin{aligned}
  m &= \frac{2\sqrt{\pi}Nd^{2}}{\hbar k\upsilon_{T}},   &\qquad
  \xi &= kx,                                           &\qquad
  \phi &= kl, \\[2pt]
  \nu &= \upsilon/\upsilon_{T},                        &\qquad
  \Gamma &= \gamma/k\upsilon_{T},                      &\qquad
  \Omega &= (\omega-\omega_{21})/k\upsilon_{T},
\end{aligned}
\label{eq:dimensionless}
\end{equation}
together with $\eta=\Gamma-i\Omega$, and
$\sigma(\xi,\nu)=(2\hbar k\upsilon_{T}/id)\,\rho_{21}(\xi,\nu)$ where $\rho_{21}$ denotes the off-diagonal density-matrix element. As in the common notation, $N$ is the
atomic density, $d$ the dipole moment of the transition, $k=2\pi/\lambda$ is the
wave number, and $\gamma$ represents the homogeneous
width of the transition. The single parameter
$m$ defines the optical density of the atomic vapor that is proportional to the atomic density $N$.

\subsection{Maxwell--Bloch equations and quenching boundary conditions}
\label{sec:mb}

Employing the semi-classical approach, the self-consistent problem for the spatial-dependent field amplitude $E(\xi)$ and the manifold of the normalised
off-diagonal density-matrix elements $\sigma(\xi,\nu)$ describing the behaviour of the atoms moving along $x$-axis with the velocity class
$\nu$, can be written in the following way using the dimensionless variables
\begin{gather}
  \frac{d^{2}E}{d\xi^{2}}+E(\xi)=-2im\,\bar\sigma(\xi),
  \qquad
  \bar\sigma(\xi)\equiv\int_{-\infty}^{\infty}\!\sigma(\xi,\nu)\,
  e^{-\nu^{2}}\,d\nu,
  \label{eq:MBfield}\\[2pt]
  \nu\,\frac{\partial\sigma(\xi,\nu)}{\partial\xi}+\eta\,\sigma(\xi,\nu)=E(\xi).
  \label{eq:MBmat}
\end{gather}
For the boundary conditions we consider here the case of quenching of the atomic polarisation at the two walls implying that an atom leaves
either surface with no polarisation after the collision,
\begin{equation}
  \sigma(0,\nu)=0 \;\;(\nu>0),
  \qquad
  \sigma(\phi,\nu)=0 \;\;(\nu<0).
  \label{eq:BCsigma}
\end{equation}
Integrating Eq.~\eqref{eq:MBmat} with these conditions
gives, for the two directions of flight,
\begin{equation}
\sigma(\xi,\nu)=
\begin{cases}
\dfrac{1}{\nu}\displaystyle\int_{0}^{\xi}E(\xi')\,
   e^{-\eta(\xi-\xi')/\nu}\,d\xi', & \nu>0,\\[2.4ex]
\dfrac{1}{\nu}\displaystyle\int_{\phi}^{\xi}E(\xi')\,
   e^{-\eta(\xi-\xi')/\nu}\,d\xi', & \nu<0.
\end{cases}
\label{eq:sigma_char}
\end{equation}
Next, averaging over the Maxwellian distribution and interchanging the $\nu$ and $\xi'$
integrations, the two half-ranges merge into a single kernel that depends only
on the separation $|\xi-\xi'|$,
\begin{equation}
  -2im\,\bar\sigma(\xi)
  =m\!\int_{0}^{\phi}\!\bar K(|\xi-\xi'|)\,E(\xi')\,d\xi',
  \label{eq:sourceK}
\end{equation}
with
\begin{equation}
  \bar K(s)=\int_{0}^{\infty}\!K(s,\nu)\,d\nu,
  \qquad
  K(s,\nu)=-\frac{2i}{\nu}\,e^{-\nu^{2}-\eta s/\nu}.
  \label{eq:kernel}
\end{equation}
Equations~\eqref{eq:MBfield} and~\eqref{eq:sourceK} combine into an
integro-differential equation for the field $E(\xi)$
alone,
\begin{equation}
  \frac{d^{2}E}{d\xi^{2}}+E(\xi)
  =m\!\int_{0}^{\phi}\!\bar K(|\xi-\xi'|)\,E(\xi')\,d\xi'.
  \label{eq:field22}
\end{equation}
Now, inverting the operator $d^{2}/d\xi^{2}+1$ by variation of constants and
anchoring the particular integral at the center of the layer, $\xi=\phi/2$,
yields~\cite[Eq.~(26)]{EV2022}
\begin{equation}
E(\xi)=C_{1}\cos\!\Big(\xi-\phi/2\Big)
         +C_{2}\sin\!\Big(\xi-\phi/2\Big) +m\!\int_{\phi/2}^{\xi}\!\sin(\xi-\xi')
  \int_{0}^{\phi}\bar K(|\xi'-\xi''|)\,E(\xi'')\,d\xi''\,d\xi'.
\label{eq:E26}
\end{equation}
The mathematical framework for boundary-value problems of this type, with
conditions imposed in different regions of velocity space, is described in detail in
Ref.~\cite{SMIRNOV}.

\subsection{Reflection and transmission}
\label{sec:observables}

Because $\bar K$ depends only on $|\xi'-\xi''|$ and the interval is symmetric
around $\phi/2$, the solution splits into even and odd eigenmodes,
$E_{e}$ and $E_{o}$ that for $m=0$ reduce to
\begin{equation}
  E^{[0]}_{e}=\cos\!\Big(\xi-\phi/2\Big),
  \qquad
  E^{[0]}_{o}=\sin\!\Big(\xi-\phi/2\Big).
  \label{eq:seeds}
\end{equation}
Using the continuity of the tangential field components at the two vapor--dielectric
interfaces we can derive the amplitude reflection and transmission
coefficients,
\begin{align}
  r&=\frac{(n_{1}-n_{2})(I_{1}I_{4}+I_{2}I_{3})
          +2i(n_{1}n_{2}I_{1}I_{2}+I_{3}I_{4})}
          {(n_{1}+n_{2})(I_{1}I_{4}+I_{2}I_{3})
          +2i(n_{1}n_{2}I_{1}I_{2}-I_{3}I_{4})},
  \label{eq:r}\\[4pt]
  t&=\frac{2n_{1}(I_{1}I_{4}-I_{2}I_{3})}
          {(n_{1}+n_{2})(I_{1}I_{4}+I_{2}I_{3})
          +2i(n_{1}n_{2}I_{1}I_{2}-I_{3}I_{4})},
  \label{eq:t}
\end{align}
where the four boundary numbers arising from the dielectric boundary conditions~\cite{Born1999}
\begin{equation}
\begin{aligned}
  I_{1}&=E_{e}|_{\xi=0}=E_{e}|_{\xi=\phi}, &\quad
  I_{2}&=E_{o}|_{\xi=0}=-E_{o}|_{\xi=\phi}, \\
  I_{3}&=\partial_{\xi}E_{e}|_{\xi=0}=-\partial_{\xi}E_{e}|_{\xi=\phi}, &\quad
  I_{4}&=\partial_{\xi}E_{o}|_{\xi=0}=\partial_{\xi}E_{o}|_{\xi=\phi}.
\end{aligned}
\label{eq:I1234}
\end{equation}
The reflectivity can be then computed as $R=|r|^2$. Note that so far no assumptions have been made for the field structure inside the vapor layer. Equations~\eqref{eq:MBfield}--\eqref{eq:I1234} are exact for the considered two-level problem in the resonance approximation and linear regime of interactions. In the following section we derive the rigorous numerical scheme for evaluating eigenmodes $E_{e}$ and $E_{o}$.

\section{Results}
\label{sec:results}

\subsection{Exact solution to all orders in the optical density}
\label{sec:exact}

We start by introducing the three linear operators:
\begin{align}
  (\hat{\mathcal K}E)(\xi')
    &=\int_{0}^{\phi}\!\bar K(|\xi'-\xi''|)\,E(\xi'')\,d\xi'',
  \label{eq:opK}\\
  (\hat{\mathcal S}g)(\xi)
    &=\int_{\phi/2}^{\xi}\!\sin(\xi-\xi')\,g(\xi')\,d\xi',
  \qquad
  \hat{\mathcal T}\equiv\hat{\mathcal S}\hat{\mathcal K}.
  \label{eq:opS}
\end{align}
Here $\hat{\mathcal K}$ performs the convolution of  the field $E(\xi)$ with the constructed nonlocal kernel $\bar{K}$ and $\hat{\mathcal S}$ is the Green's function of
$d^{2}/d\xi^{2}+1$ anchored at the center of the layer. With this notation
Eq.~\eqref{eq:E26} becomes
\begin{equation}
  E=C_{1}\cos\!\Big(\xi-\phi/2\Big)
   +C_{2}\sin\!\Big(\xi-\phi/2\Big)+m\,\hat{\mathcal T}E.
  \label{eq:operator}
\end{equation}
Reference~\cite{EV2022} solves Eq.~\eqref{eq:operator} by expanding the field in
powers of the optical density,
\begin{equation}
  E_{e/o}=E^{[0]}_{e/o}+m\hat{\mathcal T}E^{[0]}_{e/o}
         +m^{2}\hat{\mathcal T}^{2}E^{[0]}_{e/o}+\dots
         =\sum_{p\geq0}\big(m\hat{\mathcal T}\big)^{p}E^{[0]}_{e/o},
  \label{eq:series}
\end{equation}
and truncating at the desired order, which is what restricts that treatment to
low densities. The central observation of this work is that
Eq.~\eqref{eq:series} can be seen as the solution of the following integral equation
\begin{equation}
  \big(\hat 1-m\hat{\mathcal T}\big)\,E_{e/o}=E^{[0]}_{e/o},
  \label{eq:master}
\end{equation}
so that summing the expansion to all orders is the same as inverting
$\hat1-m\hat{\mathcal T}$ directly. Directly solving Eq.~\eqref{eq:master} retains every order in $m$ and covers the optically dense regime, and the intermediate
layer thicknesses, for which the expansion in $m$ should fail.

\subsection{Numerical scheme}
\label{sec:numerics}

\begin{figure}[h]
\centering
\includegraphics[width=0.5\linewidth]{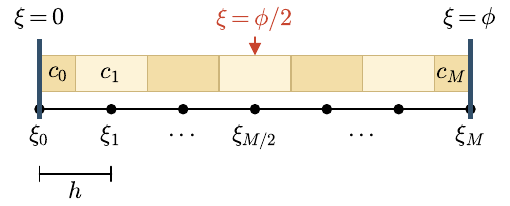}
\caption{\label{fig:discretisation} Illustration of the spatial discretisation used to construct the exact numerical scheme. The vapor layer of thickness $\phi=kl$ is split into $M+1$ cells using the uniform grid with a step $h=\phi/M$. The field is considered to be piecewise-constant in each cell allowing the evaluation of the operators $\hat{S}$ and $\hat{K}$ in a matrix form (see text for details).}
\end{figure}

On a uniform grid $\xi_{j}=jh$, $j=0,\dots,M$ with $M$ being an even number and
$h=\phi/M$ (see Fig.~\ref{fig:discretisation}), each operator becomes an $(M{+}1)\times(M{+}1)$ matrix and each
field is a complex vector of length $M+1$.

Equation~\eqref{eq:master} now can be represented as a linear system of equations~\cite{Atkinson1997} with the right-hand side defined as in Eq.~\eqref{eq:seeds}
\begin{equation}
  \big(\bm{1}-m\,\bm{S}\bm{K}\big)\,\bm{E}_{e/o}=\bm{E}^{[0]}_{e/o},
  \label{eq:linsys}
\end{equation}
which is solvable whenever $\big(\bm{1}-m\,\bm{S}\bm{K}\big)^{-1}$ exists. The matrix $\bm{S}$ has a smooth kernel and is discretised here by the
trapezoidal rule. For the kernel matrix $\bm{K}$ the singularity of
$\bar K$ as $s\to0$ is integrated analytically over each cell. Specifically, taking the field
piecewise-constant in each cell $c_0$--$c_M$, which is the only approximation, the required cell integrals are differences of
\begin{equation}
  G(a)=\int_{0}^{a}\!\bar K(s)\,ds
      =-\frac{i \sqrt{\pi}}{\eta}+\frac{2i}{\eta}J(\eta a),
  \qquad
  J(b)=\!\int_{0}^{\infty}\!e^{-\nu^{2}-b/\nu}\,d\nu,
  \label{eq:G}
\end{equation}
so the singular point never enters pointwise~\cite{Trefethen2014}. The numerical solution of Eq.~\eqref{eq:linsys} yields $\bm{E}_{e}$ and $\bm{E}_{o}$ at each detuning. The latter are then inserted into Eq.~\eqref{eq:I1234}, which allows one to calculate the amplitudes from Eqs.~\eqref{eq:r}--\eqref{eq:t}. Convergence is $O(h^{2})$, and a resolution
$M\gtrsim2^5\,l/\lambda$ is sufficient for most use cases. For the details of the numerical simulation scheme and the particular programming implementation, we address readers to Data Availability statement.

Throughout Sec.~\ref{sec:results} the first-order PT curves shown for comparison
are computed from the same matrices $\bm{S}$ and $\bm{K}$, as
$\bm{E}^{[1]}=\bm{E}^{[0]}+m\bm{S}\bm{K}\bm{E}^{[0]}$ with the results truncated consistently at $O(m)$. Differences between the exact and the
first-order curves are therefore higher-order physics in $m$, not discretization
error.

\subsection{First-order non-selective and selective contributions}
\label{sec:firstorder}

In what follows we derive an approximate analytical expression for the optimal nanocell thickness that maximises the absolute amplitude of the first derivative of the reflectivity $\partial R / \partial \Omega$ that is commonly measured in experiments. We begin by separating the total reflection from the vapor cell into non-selective (reflection from the cell windows) and selective (reflection from the vapor) contributions. In the case of identical surrounding dielectric media considered here ($n_{1}=n_{2}$), the window--vapor interface reflects with amplitude $r_0{}=(n_{1}-1)/(n_{1}+1)$
on the entrance side and $-r_0{}$ on the exit side, and the two contributions have the round-trip phase difference of $2\phi$. Denoting by $E_{0}$ the amplitude of the incident field (see Fig.~\ref{fig:geometry}), to lowest order in $r_0$ this yields the following contribution,
\begin{equation}
  \frac{E_{r0}}{E_{0}}=r_0{}\big(1-e^{2i\phi}\big)
   =-2ir_0{}\sin\phi\;e^{i\phi},
  \label{eq:background}
\end{equation}
\noindent which next mixes up with the atomic response. For the sake of clarity here we temporarily omit
the full FP response.

In the first-order PT in vapor density, and in the sub-Doppler window $\Omega\sim\Gamma\ll1$,
$\phi\ll\Gamma^{-1}$, there exists an analytical expression for 
the field selectively reflected by a thin vapor layer (see Eqs.~(15) and~(16) of Ref.~\cite{PhysRevA.51.1959}). Retaining
only the terms that depend on the detuning and combining all the rest into a constant independent of the detuning, it reads as
\begin{equation}
  \frac{E_{r}(\Omega)}{E_{0}}
   =2m\,(1-\cos\phi)\,e^{i\phi}\,\ln(\Gamma-i\Omega)+\mathrm{const}.
  \label{eq:compact}
\end{equation}
\noindent Here we use Eq.~\eqref{eq:compact}
directly, while Appendix~\ref{app:log} shows how the complex logarithm arises from the nonlocal integral kernel.
Adding Eqs.~\eqref{eq:compact} and~\eqref{eq:background} and normalising
to the incident intensity yields the reflectivity [here we used $\ln(\Gamma-i\Omega) = 1/2\ln(\Gamma^2+\Omega^2) - i\arctan(\Omega/\Gamma)$]
\begin{equation}
  R(\Omega)
  =\underbrace{4r_0{}^{2}\sin^{2}\phi}_{\text{empty cell}}
  \;+\;\underbrace{8r_0{}\,m\,(1-\cos\phi)\sin\phi\,
        \arctan(\Omega/\Gamma)}_{\mathcal{O}(m):\ \text{odd spectral line contour}}
  \;+\;\mathcal{O}(m^2),
\label{eq:lineshape}
\end{equation}
up to an $\Omega$-independent $\mathcal{O}(m)$ term. Equation~\eqref{eq:lineshape} shows how the spectral line contour of reflection looks for a symmetric cell with two uncoated windows.

\begin{enumerate}
\item The term linear in the vapor density is proportional to
      $\arctan(\Omega/\Gamma)$ and is therefore odd in the detuning
      for every thickness. Once the common factor $e^{i\phi}$ is removed, the
      background is purely imaginary, the even contour is purely real and
      the odd contour is purely imaginary. The interference cross term thus
      selects $\arctan(\Omega/\Gamma)$ and annihilates $\ln(\Gamma^2+\Omega^2)$, whatever $\phi$ may be.

\item Consequently, to the first order in $m$, the derivative of the reflectivity is
      \begin{equation}
        \frac{\partial R}{\partial\Omega}
        =8r_0{}m(1-\cos\phi)\sin\phi\;
         \frac{\Gamma}{\Gamma^{2}+\Omega^{2}}+\mathcal{O}(m^{2}).
        \label{eq:derivative}
      \end{equation}

\item The even (logarithmic) contour survives only at second order in $m$~\cite{Dutier:03}. It becomes the leading behavior where the odd term vanishes, specifically when
      $\sin\phi=0$, but only at the odd multiples of $\pi$. At
      $\phi=(2n+1)\pi$ or equivalently $l=(2n+1)\lambda/2$, where $n=0,1,2,...$, one has
      $\cos\phi=-1$, the non-selective background vanishes and the $\mathcal{O}(m^{2})$
      contribution is all that remains. At the even multiples,
      $\phi=2n\pi$ ($l=n\lambda$), both $\sin\phi=0$ and
      $\cos\phi=1$, so that the common factor $(1-\cos\phi)$ annihilates
      the linear and quadratic terms simultaneously, so that only the Doppler-broadened signal can be observed.

\end{enumerate}

\subsection{Optimal layer thickness for the first derivative of the reflection spectrum}
\label{sec:thickness}

The amplitude of the first-order derivative $\partial R/\partial \Omega$ in Eq.~\eqref{eq:derivative} is governed by the following term $(1-\cos\phi)\sin\phi$, whose stationary points are 
$\cos\phi=-1/2$ and $\cos\phi=1$. First, stationary point $\cos\phi=1$ leads to a zero amplitude of the first derivative of the reflection spectrum with respect to detuning. Therefore, this condition is not particularly interesting for us. Contrary to that, the condition $\cos \phi = -1/2$ maximises the amplitude of the odd spectral line contour of the reflectivity (see Eq.~\eqref{eq:lineshape}) and consequently maximises the peak absolute amplitude of the even spectral line contour of the first derivative of the reflectivity with respect to the detuning (see Eq.~\eqref{eq:derivative}). The optimal thickness, thus, yields two branches of solutions
\begin{equation}
  \phi_{\mathrm{opt}}=2\pi/3+2\pi n;\,\,4\pi/3+2\pi n,
  \quad
  l_{\mathrm{opt}}=\lambda/3+n\lambda;\,\, 2\lambda/3+n\lambda,
  \label{eq:opt}
\end{equation}
with the sign of the first derivative peak flipping at every $\phi=n\pi$ (or similarly $l=n\lambda/2$).

\begin{figure}[t!]
\centering
\includegraphics[width=0.5\linewidth]{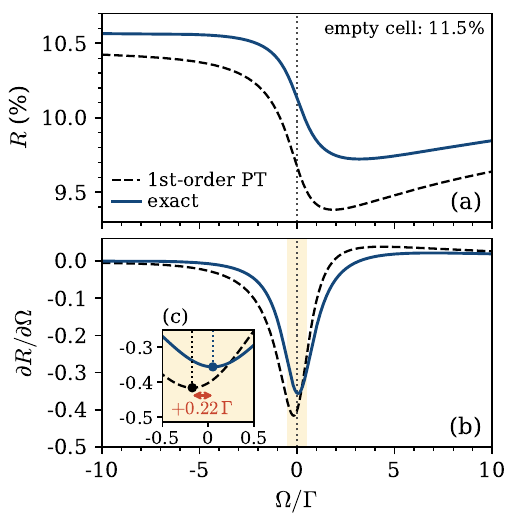}
\caption{\label{fig:l5over3} (a) Reflection spectrum from the exact, all-order solution of
Eq.~\eqref{eq:master} (blue solid) and from first-order PT, with $R$ truncated at
first order in $m$ (black dashed), for
$l=5\lambda/3$ ($\phi=10\pi/3$), $m=0.002$, $\Gamma=0.01$ and
$n_{1}=n_{2}=1.5$. The empty-cell FP background reflection level is given in the
legend. (b) The corresponding first derivative
$\partial R/\partial\Omega$ on the same detuning axis. In (a) and (b) the dotted
vertical line marks $\Omega=0$. (c) Zoom into the derivative of reflection spectra near resonance. On this scale
the density-dependent shift of the peak between the exact solution and
first-order theory is resolved and the dotted vertical lines
mark the two peak positions.}
\end{figure}

As an illustrative example, below we consider the case $l=5\lambda/3$ that
satisfies the condition $\cos\phi=-1/2$ exactly. Figure~\ref{fig:l5over3} shows the reflection and corresponding first derivative spectra for the calculations performed in the framework of the first-order PT (black dashed curve) and using the introduced rigorous numerical scheme (blue solid curve). We can see that reflectivity shows approximately odd spectral line contour resulting in approximately even spectral line contour with one dip for the first derivative spectrum [see Fig.~\ref{fig:l5over3}]. 

Equation~\eqref{eq:opt} can provide guidance for choosing an appropriate nanocell thickness in experiments. For example, the common choice of an odd multiple of $\lambda/4$ that satisfies $\cos\phi=0$ and leads to the highest FP background reflection level, however, gives central amplitude $\max(|\partial R/\partial\Omega|)$ of $\sim 0.2$ at $l=7\lambda/4$ (see Fig.~\ref{fig:l7over4} in Appendix B),
against $\sim 0.36$ at the nearest $l=5\lambda/3$ optimum when solved exactly for $n_{1}=n_{2}=1.5$, at $m=0.002$ and $\Gamma=0.01$. Here we want to stress that for the sake of simplicity, Eq.~\eqref{eq:opt} was derived analytically in the first order with respect to the atomic density $m$, but also in the first order with respect to the window reflectivity $r_0$. Therefore, we expect some deviations of optimal thickness at larger layer thicknesses, higher densities and window reflectivities, which will be discussed in the following section.

\begin{figure}[t!]
\centering
\includegraphics[width=0.5\linewidth]{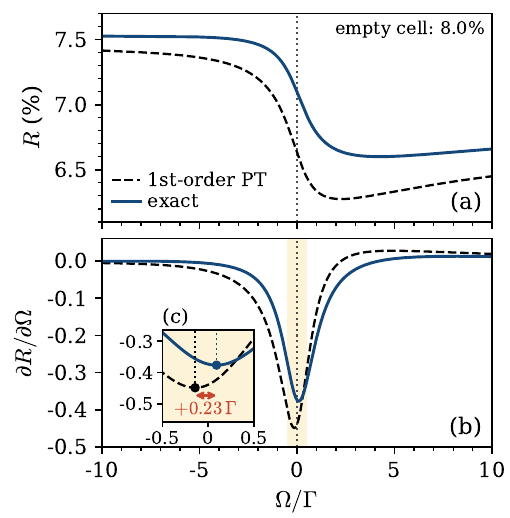}
\caption{\label{fig:l13over8}(a) Reflection spectrum from the exact, all-order solution of
Eq.~\eqref{eq:master} (blue solid) and from first-order PT, with $R$ truncated at
first order in $m$ (black dashed), for
$l=13\lambda/8$ ($\phi=13\pi/4$), $m=0.002$, $\Gamma=0.01$ and
$n_{1}=n_{2}=1.5$. The empty-cell FP background reflection level is given in the
legend. (b) The corresponding first derivative
$\partial R/\partial\Omega$ on the same detuning axis. In (a) and (b) the dotted
vertical line marks $\Omega=0$. (c) Zoom into the derivative of reflection spectra near resonance. On this scale
the density-dependent shift of the peak between the exact solution and
first-order theory is resolved and the dotted vertical lines
mark the two peak positions.}
\end{figure}

In Fig.~\ref{fig:l13over8}, we show the spectra obtained for the layer thickness $l=13\lambda/8$ that is slightly shifted from the $l=5\lambda/3$ optimum. Despite the fact that the considered vapor layer thickness is suboptimal according to Eq.~\eqref{eq:opt}, the spectrum of the first derivative gives slightly higher absolute peak amplitude, yet at the expense of the lower FP background reflection level. This also emphasises what is lost when the problem is treated
perturbatively. The aforementioned limitation will be partially resolved in the following section.

We can also notice that the spectra computed using exact and the first-order PT approaches differ not only in
amplitude, but also in the position of the resonance [see Figs.~\ref{fig:l5over3}(c) and~\ref{fig:l13over8} (c)]. The peak of
$|\partial R/\partial\Omega|$ is shifted towards the blue side of the spectrum by approximately
$0.22\,\Gamma$ and $0.23\,\Gamma$ at $l=5\lambda/3$ and $l=13\lambda/8$, respectively, when compared to the first-order PT results. At nearly a quarter of $\gamma$ it is comparable
to the linewidths well resolved in sub-micron cells, so that a
perturbative analysis would misplace the resonance significantly. This is the density-dependent shift first found in semi-infinite vapor layers~\cite{schuurmans:jpa-00208442,GUO1996,BLUESHIFT} (see also Appendix B, Fig.~\ref{fig:l500} for details). In Ref.~\cite{EV2022} this so-called blue shift was studied using absorption spectra of a thin vapor layer for the case of quenching atom-wall collisions considered in this work. Its values were shown to grow approximately linearly with atomic density $m$ and depend substantially on the layer thickness $\phi$. In the case of reflection spectra studied here, the situation is more complex, as the higher order density effects also result in the thickness-dependent line shape distortion of $R(\Omega)$. The proposed numerical scheme fully captures these higher-order density effects.

\subsection{Accurate expression for the optimal thickness}
\label{sec:optimal}

\begin{figure*}[b!]
\centering
\includegraphics[width=\linewidth]{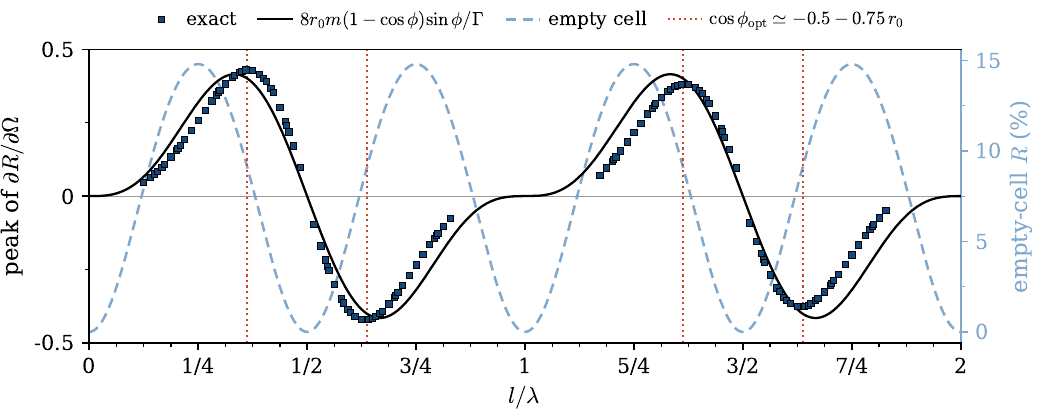}
\caption{\label{fig:thickness}Peak value of $\partial R/\partial\Omega$ as a
function of the layer thickness: numerical calculation (blue squares)
versus the first-order analytical prediction (black curve, Eq.~\eqref{eq:derivative}),
see text for details. The dashed curve (right axis) is the FP background reflection level as a
function of the layer thickness. Dotted vertical lines mark the optimal thickness
estimated from $\cos\phi\simeq-0.5-0.75r_0$.
The following parameters are used in simulations: $m=0.002$, $\Gamma=0.01$ and $n_{1}=n_{2}=1.5$.}
\end{figure*}

In Fig.~\ref{fig:thickness} using the results of the rigorous numerical modeling and analytical analysis performed in the previous sections, we show how various properties of nanocell reflection spectra depend on the vapor layer thickness which we vary here in the range from $l=\lambda/8$ up to $2\lambda$. First of all, the dashed curve in Fig.~\ref{fig:thickness} (right axis) shows the background FP reflection level of an empty cell with two identical windows ($n_1=n_2=1.5$), manifesting the well-known $\lambda/2$-periodic dependence on the layer thickness. Next, the blue squares in Fig.~\ref{fig:thickness} represent the amplitude of the peak of $\partial R / \partial \Omega$ obtained using the described rigorous numerical model. We note here that the numerical simulations in the vicinities of $l=n \lambda/2$ are not shown in Fig.~\ref{fig:thickness} for the following reasons. First of all, as mentioned above, the given condition corresponds to the condition resulting in the absence of the atomic contribution in the first order in $m$, so the signal is strongly suppressed and, thus, would be very hard to detect experimentally. Moreover, in the vicinity of these thicknesses the first-order derivative of the reflectivity $\partial R/\partial \Omega$ shows approximately odd spectral line contour that does not allow one to extract a single peak amplitude value. The numerical scheme, however, can be applied for any thickness of the vapor layer. 

Next, the black solid curve plots analytical Eq.~\eqref{eq:derivative} truncated at the first order in $m$. We can see that the analytical result shows almost perfect agreement with the rigorous numerical modeling for thin vapor cells ($l\lesssim\lambda$) and well captures the $\lambda$-periodic dependence of the atomic contribution with respect to the layer thickness~\cite{PhysRevA.51.1959}. As expected, the analytical expression starts to diverge from the numerical simulations for thicker cells. In particular, one can see how the optimal vapor layer thickness shifts due to higher-order effects neglected in Eq.~\eqref{eq:derivative}. Indeed, Eq.~\eqref{eq:background}, which we used to derive an analytical expression for the optimal vapor layer thickness, keeps only the two lowest reflection paths and
adds the first-order PT field solution directly to them. A real cell is a three-layer stack, and
two contributions are missing in the analytical treatment presented in Sec.~\ref{sec:firstorder}, namely, the multiple reflections
between the windows and the field transmitted through the
layer and returned by the rear window. Once they are included,
the position of the optimum becomes a function of $r_0$, which we approximate here up to the first order
in $r_0$ by fitting to the numerical optima
\begin{equation}
  \cos\phi_{\rm opt}\simeq-0.5-0.75\,r_0.
  \label{eq:expansion}
\end{equation}
Equation~\eqref{eq:expansion} has been tested numerically for symmetric cells
with refractive indices in the range $n_{1}=n_{2}\in[1.1,\,2.5]$ and reproduces
the layer thickness that maximises the amplitude of the first-derivative
spectrum almost perfectly and far below the uncertainties usually encountered in nanocell
fabrication. For the case shown in Fig.~\ref{fig:thickness} ($n_1=n_2=1.5$, i.e. $r_0=0.2$), Eq.~\eqref{eq:expansion} gives $\cos\phi_{opt} \approx -0.65$ and the corresponding optimal vapor layer thicknesses are shown by the dotted vertical lines. Specifically, it gives the fourth optimum near $l\approx1.64 \lambda$, which is very close to the earlier considered case $l=13\lambda/8$ (see Fig.~\ref{fig:l13over8}).

\section{Discussion \& Conclusion}
\label{sec:discussion}

In this work we proposed an exact numerical simulation scheme that reduces the
self-consistent set of Maxwell--Bloch equations for the field and the atomic
polarisation to a linear system of equations that we solve by direct matrix
inversion. Unlike the earlier PT treatment in optical density~\cite{EV2022}, our solution is
accurate in vapor density and holds for any layer
thickness. This allowed us to solve two important problems associated with the
linear interaction of resonant radiation with a thin vapor layer. First, we
determined the layer thickness that maximises the amplitude of the first
derivative of the reflectivity with respect to the laser detuning.
Specifically, we derived a simple analytical expression for the optimal
thicknesses, $l=\lambda/3+n\lambda$ and $2\lambda/3+n\lambda$ with $n=0,1,2,...$, valid to first order in the atomic density and in the
window reflectivity $r_0$. We then extended this result with the help of the numerical scheme and obtained a simple formula [see Eq.~\eqref{eq:expansion}] which reproduces the exact optima for
symmetric cells with sufficient accuracy. These expressions can be used to
guide the design of vapor nanocells in experiments.

Second, our numerical scheme fully captures the shift and distortion of the
spectral line contour originating from the higher-order density effects and absent in the first-order PT. We demonstrated that the shift of the peak of
$\partial R/\partial\Omega$ can become comparable to homogeneous
width of the spectra line. Shifts of this size matter in practice. Derivative
selective-reflection spectra are routinely used to locate atomic transitions
with sub-Doppler precision and to compare them with model line
shapes~\cite{DucloyFichet1991,Sargsyan2026}. When the aim is to measure the
atom--surface interaction, i.e.\ the van der Waals or Casimir--Polder shift,
the interaction constant is obtained by attributing the measured shift and
distortion of the line to the surface
potential~\cite{BlochDucloy2005,Fichet2007,Laliotis2014,ADAMS2}.
Density-dependent shifts of a comparable size, including collective
shifts~\cite{KEAVENEY,GUO1996}, must then be modelled and separated from the
surface contribution.

This work also takes another step toward the microscopic theory of the optical
response of confined atomic vapors. Unlike the conventional dispersion theory,
which describes the vapor by a local susceptibility, it represents the
atom--wall collisions by the boundary conditions imposed on the atomic
polarisation of each velocity class [see Eq.~\eqref{eq:BCsigma}] and thereby
retains the transient response of atoms leaving the walls, which makes the
optical response of the layer nonlocal~\cite{ADAMS}. Since different boundary
conditions lead to different sub-Doppler spectra, selective reflection from
thin layers probes the atom--wall collisions themselves, and the same transient
response must be taken into account when atom--surface interactions are
extracted from nanocell spectra~\cite{Bloch2015,Whittaker2014}.

Finally, we would like to outline the future directions of this work. First,
the exact expressions for the optimal thicknesses can, of course, be extended in validity to all orders in $r_0$ and the case of non-symmetric windows (i.e. $n_1 \neq n_2$), which was not done
here for the sake of clarity. Second, the van der Waals atom--surface
interactions~\cite{DucloyFichet1991,Fichet2007} and the Lorentz--Lorenz
local-field correction~\cite{GUO1996} can be included by modifying only the
kernel matrix $\bm{K}$ in Eq.~\eqref{eq:linsys}. Importantly, the even and odd
eigenmodes of the vapor layer do not depend on the surrounding media, which
enter only through Eqs.~\eqref{eq:r}--\eqref{eq:t}, so the same scheme can be
used to explore various nanocell geometries, for example, wedge-shaped
cells~\cite{SARKISYAN2001} or various microfabricated
vapor cells~\cite{Klinger2025, Peroux2025}. Extending the proposed method for these applications will
allow a consistent treatment of surface-induced and density-dependent shifts in
the spectra of vapor nanocells.

\newpage

\section*{funding}
The authors received no external funding for this work.

\section*{Data availability}
\label{sec:code}
The program code used to perform the rigorous numerical simulations shown in this work is publicly available at \url{https://github.com/AndreiErmolaev/TVL-spectra}. All the other data shown in this work can be sent by authors upon reasonable request.

\section*{Competing interests}
We declare that none of the authors have competing financial or non-financial interests.

\newpage

\appendix
\section{First-order contribution of atomic vapor to the reflected field}
\label{app:log}

This appendix derives the detuning-dependent part of the reflected field in the first-order PT in optical density. 
At first order in $m$, the detuning appears only
in $\eta=\Gamma-i\Omega$ inside the
nonlocal kernel $\bar K(s)$ [see Eq.~\eqref{eq:kernel}]. The kernel can be directly
expressed through the derivative of the special function
$J(b)=\int_{0}^{\infty}e^{-\nu^{2}-b/\nu}\,d\nu$ [see Eq.~\eqref{eq:G}], where $b=\eta s$. We first differentiate $J(b)$ with respect to $b$ and then compare the result with
Eq.~\eqref{eq:kernel}, this yields
\begin{equation}
 \frac{dJ}{db}=-\int_{0}^{\infty}\frac{d\nu}{\nu}\,e^{-\nu^{2}-b/\nu}, \qquad
  \bar K(s)=2i\,\frac{dJ}{db},
  \label{eq:dI0}
\end{equation}
\noindent Restricting our consideration to the case of small detunings (i.e. $\Omega \sim \Gamma\ll1$) and optically-thin vapor layers $\phi\ll\Gamma^{-1}$ and taking into account that $0\le s\le\phi$, we notice that the argument $b$ remains small $|b|\le\phi|\eta|=\phi\sqrt{\Gamma^2+\Omega^2} \ll1$. Next, we can see that the factor $e^{-b/\nu}$ cuts the integral off below at
$\nu\sim|b|$, and the Gaussian cuts it off above at $\nu\sim1$. Between these
two limits the integrand is simply $1/\nu$ and its integration is straightforward:
\begin{equation}
  \frac{dJ}{db}\approx\ln b+\mathrm{const},
  \label{eq:dI0log}
\end{equation}
and gives the complex logarithm form of the solution. Because $\ln b=\ln\eta+\ln s$, the kernel splits into two parts,
\begin{equation}
  \bar K(s)\approx 2i\ln\eta+2i\left(\ln s+\mathrm{const}\right),
  \label{eq:Klog}
\end{equation}
of which only the first depends on the detuning, and it does not depend on $s$. Now, inserting Eq.~\eqref{eq:Klog} into Eq.~\eqref{eq:field22}, the field equation
simplifies to
\begin{equation}
  \frac{d^{2}E}{d\xi^{2}}+E(\xi)\approx
  2im\ln\eta\int_{0}^{\phi}E(\xi')\,d\xi',
  \label{eq:fieldlog}
\end{equation}
up to the term independent of detuning. The entire detuning dependence is thus carried by a source term that is proportional to
$\ln\eta=\ln(\Gamma-i\Omega)$. In the framework of the first-order PT, we assume that the driving field is unperturbed and neglect the windows. This gives $E(\xi)= E_0e^{i\xi}$ inside the layer and the solution of Eq.~\eqref{eq:fieldlog} for the reflected field can be written in terms of Green's function (see Ref.~\cite{PhysRevA.51.1959}) as
\begin{equation}
  \dfrac{E_r(\Omega)}{E_0}\approx m\ln\eta\int_0^\phi\int_{0}^{\phi}e^{i\xi}e^{i\xi'}\,d\xi d\xi'
  =2m\,(1-\cos\phi)\,e^{i\phi}\ln\eta,
  \label{eq:geom}
\end{equation}
which is
Eq.~\eqref{eq:compact} up to a detuning-independent constant.

\section{Spectra at the conventional quarter-wave thickness and in a thick cell}
\label{app:extra}

To further illustrate the range of the proposed numerical scheme, below we consider two more examples both of which are
computed for the same parameters as Figs.~\ref{fig:l5over3}
and~\ref{fig:l13over8}, namely $m=0.002$, $\Gamma=0.01$ and $n_{1}=n_{2}=1.5$, but for different layer thicknesses $l$.

\begin{figure}[ht!]
\centering
\includegraphics[width=0.5\linewidth]{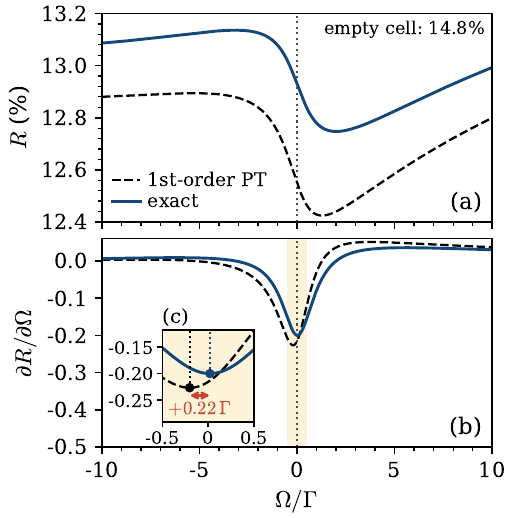}
\caption{\label{fig:l7over4}As Fig.~\ref{fig:l5over3}, for the conventional
choice $l=7\lambda/4$ ($\phi=7\pi/2$, $\cos\phi=0$). The peak absolute amplitude
$\partial R/\partial\Omega$ is almost two times smaller as opposed to the cases shown in Figs.~\ref{fig:l5over3}-\ref{fig:l13over8}.}
\end{figure}

Figure~\ref{fig:l7over4} shows the case $l=7\lambda/4$, an odd multiple of
$\lambda/4$ and therefore the choice most often made in
experiments~\cite{PhysRevA.51.1959,DutierEPL2003,Sargsyan2026}. It satisfies condition
$\cos\phi=0$, which maximises the non-selective FP background, however, as was shown in this work, does not correspond to the optimal case for selective reflection contribution. Specifically, the computed $\partial R/\partial\Omega$ shows approximately even spectral line contour as before, yet, its central absolute amplitude reaches
only $\sim 0.2$, which is almost two times smaller as opposed to the nearest optimum. The difference between the exact numerical scheme and the first-order PT is quite noticeable too. As highlighted above, for thicker cells the shift and distortion of the spectral line contour become more and more pronounced.

\begin{figure}[ht!]
\centering
\includegraphics[width=0.5\linewidth]{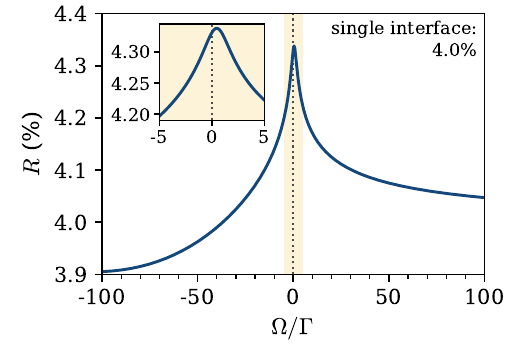}
\caption{\label{fig:l500}Reflection spectrum of a thick cell, $l=500\lambda$,
computed with the same scheme on a grid of $M=16000$ points. The inset enlarges
the shaded band $|\Omega|\le5\Gamma$. The dotted vertical line marks the position of zero detuning $\Omega=0$.}
\end{figure}

In this regard, Fig.~\ref{fig:l500} shows the reflection spectrum computed in the limit of optically long vapor cell. Here we consider a particular case $l=500\lambda$, however, at this point, the spectra practically do not change with the increase of the vapor layer thickness. Resolving the
optical oscillation over such a distance requires a substantially larger number of grid points, so we use
$M=16000$ in our simulations. The
first-order thin-layer PT analysis would predict no selective reflection in that case, as $l=500\lambda$ corresponds to the condition $l=n\lambda$ (see Sec.~\ref{sec:firstorder}). The exact solution, nevertheless, yields the well-known spectral line contour of selective reflection~\cite{Cojan1954,schuurmans:jpa-00208442}. This example shows that the scheme remains
quantitative in a regime where the PT with respect to the atomic density has no validity.

\bibliography{Ermolaev_2025_TVL_refs.bib}

\end{document}